# Low-spatial-coherence broadband fiber source for speckle free imaging

Brandon Redding[1], Peyman Ahmadi[2], Vadim Mokan[2], Martin Seifert[2], Michael A. Choma[1,3,4,5], Hui Cao[1*]

[1]Department of Applied Physics, Yale University, New Haven, Connecticut 06520, USA
[2]Nufern, East Granby, Connecticut 06026, USA
[3]Department of Diagnostic Radiology, Yale School of Medicine, New Haven, Connecticut 06520, USA
[4]Department of Pediatrics, Yale School of Medicine, New Haven, Connecticut 06520, USA
[5]Department of Biomedical Engineering, Yale University, New Haven, Connecticut 06520, USA

*Corresponding author: hui.cao@yale.edu

**We designed and demonstrate a fiber-based amplified spontaneous emission (ASE) source with low spatial coherence, low temporal coherence, and high power per mode. ASE is produced by optically pumping a large gain core multimode fiber while minimizing optical feedback to avoid lasing. The fiber ASE source provides 270 mW of continuous wave emission, centered at $\lambda$=1055 nm with a full-width half-maximum bandwidth of 74 nm. The emission is distributed among as many as ~70 spatial modes, enabling efficient speckle suppression when combined with spectral compounding. Finally, we demonstrate speckle-free full field imaging using the fiber ASE source. The fiber ASE source provides a unique combination of high power per mode with both low spatial and low temporal coherence, making it an ideal source for full-field imaging and ranging applications.**

Traditional amplified spontaneous emission (ASE) based light sources combine broadband emission, similar to a light emitting diode (LED), with high spatial coherence and high power per mode, similar to a laser. This combination, realized in both fiber-based ASE sources [1] and semiconductor-based superluminescent diodes [2], has made ASE sources increasingly popular for a range of applications including spectroscopy [3], optical coherence tomography (OCT) [4], fiber sensors [5], and gyroscopes [6]. However, the high spatial coherence of existing ASE sources has precluded their use in full-field imaging applications, where spatial coherence introduces artifacts such as speckle. By comparison, traditional low spatial coherence sources such as thermal sources and LEDs do not provide the required power per mode for high speed, full-field imaging applications [7]. Recently, there have been several demonstrations of multimode lasers which combine low spatial coherence with high power per mode, including dye-based random lasers [8,9], powder-based random Raman lasers [10], solid-state degenerate lasers [11,12], semiconductor-based chaotic microcavity lasers [13], and semiconductor-based large-area VCSELs [14–16] and VCSEL arrays [17]. However, an optical fiber based light source with low spatial coherence has not been demonstrated. In addition, each of these previous demonstrations of low spatial coherence lasers provided narrow bandwidth emission with relatively high temporal coherence, precluding their use in ranging applications such as OCT [18] or frequency resolved LiDAR [19].

In this Letter, we present a novel fiber-based ASE source that combines low temporal and low spatial coherence, similar to an LED, with the high power per spatial mode associated with lasers and traditional single spatial mode ASE sources. The fiber ASE source provides 270 mW of CW emission centered at $\lambda$=1055 nm with 74 nm 3dB bandwidth (full width at half maximum). The emission is distributed among as many as 70 spatial modes, enabling efficient speckle suppression when combined with spectral compounding. We also demonstrate speckle-free full field imaging using the fiber ASE source. By providing broadband, speckle free emission with ~40dB higher power per mode than an LED, the fiber ASE source is ideally suited for high-speed, full-field imaging and coherent ranging applications.

To achieve highly multimode emission from an optical fiber we used a recently developed optical fiber with an extra-large mode area (XLMA) gain core [20]. The basis of the XLMA design is synthetic fused bulk silica doped with ytterbium and other co-dopants that form the active core of the fiber. This rare earth doped bulk silica is commercialized by Heraeus Quarzglas (Kleinostheim, Germany). The XLMA fiber used in our ASE source has a 100μm-diameter, Yb-doped core with numerical aperture (NA) of 0.1 that is surrounded by a 400 μm diameter octagonal inner cladding, and a 480 μm diameter outer cladding (Nufern XLMA-YTF-95/400/480). A ytterbium concentration of 1000 ppm in the core results in 7.8 dB/m absorption at 972 nm. The inner cladding provides guiding for the pump light, thus it is also called the pump core in contrast to the gain core that guides the emitted light. A schematic of the XLMA fiber cross section is shown in Fig. 1(a). We estimated the number of transverse spatial modes supported in the Yb-doped gain core of the fiber to be $M = 16 R^2 (NA)^2 / \lambda^2 = 360$ at $\lambda$=1050 nm, where $R$ is the core radius [21]. The XLMA fiber was optically pumped using two 20W laser diodes operating at $\lambda$=915 nm. A fiber combiner was used to couple the output beams from the two pump diodes into the pump core of a 4.4 meter long piece of XLMA fiber, as shown schematically in Fig. 1(b). The end of the fiber was cleaved at an angle of 4° to minimize feedback which could lead to lasing. Minimization of feedback ensured that the fiber operated as a broadband amplified spontaneous emission source with low temporal coherence [22]. Emission from the end of the XLMA was then collimated and the remaining 915 nm pump light was filtered out by a dichroic filter.

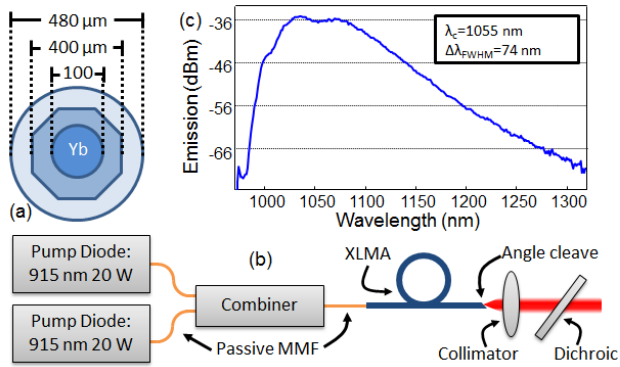

Fig. 1 (a) Cross section of the Yb-doped XLMA fiber. The fiber consists of a 100μm-diameter, Yb-doped gain core with NA=0.1 surrounded by an octagonal 400μm pump core and a 480μm outer cladding. (b) Schematic of the ASE source. Two 20W pump diodes are coupled into the pump core of the XLMA fiber, the end of which is angle-cleaved to minimize feedback. (c) The emission spectrum from the fiber ASE source is centered at 1055 nm with a 3 dB bandwidth of 74 nm.

We then characterized the emission of the fiber ASE source using a power meter and an optical spectrum analyzer. The fiber ASE source produced 270 mW of CW emission with a center wavelength of 1055 nm and a 3dB bandwidth of 74 nm, as shown in Fig. 1(c). While the increase of the emission power with the pump power is faster than linear, we did not observe saturation in the output power at the maximum pump power of 40 W, indicating that higher emission should be possible by incorporating additional pump diodes. The relatively low quantum efficiency of the current fiber ASE source is due to a mode mismatch between the passive multimode fiber of the combiner and the octagonal pump core of the XLMA fiber, which significantly reduced the amount of pump light coupled into the XLMA fiber. In addition, approximately half of the fiber ASE was in the counter-propagating direction of the pump light and not collected in our experiments. Nonetheless, the 270 mW emission in the co-propagating direction is sufficient for many imaging applications and allowed us to characterize the spatial and temporal coherence of the XLMA fiber ASE source. Moreover, the fiber ASE source already provides ~4 mW/nm, which is comparable to commercially available supercontinuum sources [23].

We then characterized the ability of the fiber ASE source to suppress speckle formation. Speckle is a coherent artifact known to corrupt image formation and can be characterized by the speckle contrast $C=\sigma_I/<I>$, where $\sigma_I$ is the standard deviation of the intensity and $<I>$ is the average intensity [24]. A recent study on the human perception of speckle found that speckle with contrast below ~0.04 could not be detected, providing a guideline for the development of a light source with sufficiently low spatial coherence for imaging [25].

To measure the speckle pattern formed by a light source, we collimated the emission onto a ground glass diffuser and recorded images of the transmitted light on a CCD camera (Allied Vision Mako-G125B). For comparison, we first measured the speckle pattern formed by light from one of the 915 nm pump diodes. As shown in Fig. 2(a), the spatially coherent 915 nm pump diode produced a clear speckle pattern with contrast of ~0.46. The speckle contrast was less than unity since the pump diode consists of a few separate emitters coupled into a multimode fiber. We then repeated the experiment while illuminating the diffuser with emission from the fiber ASE source which produced the image shown in Fig. 2(b). The uniform intensity across the image confirmed that the fiber ASE source efficiently suppressed speckle formation. Based on the image in Fig. 2(b), we calculated a speckle contrast of ~0.02, below the threshold required to avoid human detection in an imaging setting.

In addition to the 105 μm diameter XLMA fiber, we tested speckle formation using two additional ASE sources: a fiber ASE source based on a 30 μm diameter, Yb-doped fiber (Nufern LMA-YDF-30/400-VIII, NA=0.06; East Granby, CT, USA) that supports ~10 spatial modes at $\lambda$=1050 nm [21], and a commercially available, semiconductor-based multimode superluminescent diode (Superlum M-381). The 30 μm diameter fiber ASE source produced moderately broadband emission with a 3dB bandwidth of ~20 nm; however, as shown in Fig. 2(c), the emission also produced speckle with contrast of ~0.42. The multimode SLD provided ~150 mW of power at $\lambda$=800 nm with a 3dB bandwidth of 40 nm. Nonetheless, emission from the SLD still produced speckle with contrast of ~0.2. Thus, the XLMA fiber ASE source was the only ASE source that has been realized so far which suppressed speckle to acceptable levels for full-field imaging applications.

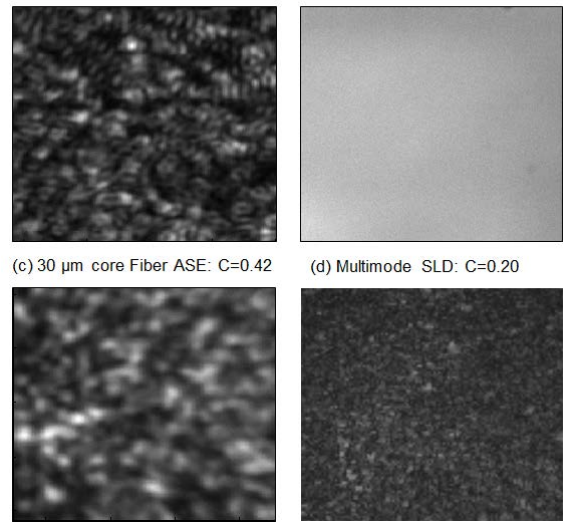

Fig. 2 Speckle formed by light incident on a CCD camera after passing through a ground glass diffuser and a linear polarizer. (a) The 915 nm pump diode used to pump the fiber ASE source produced speckle with contrast of 0.46. (b) The 100 μm diameter gain core, XLMA fiber ASE efficiently suppressed speckle, with a measured contrast of 0.02. (c) ASE from a fiber with a 30 μm diameter gain core produced speckle with contrast of 0.42. (d) ASE from a multimode SLD produced speckle with contrast of 0.2. The XLMA fiber ASE source was the only ASE source which effectively suppressed speckle formation.

The speckle contrast depends on the number of mutually incoherent spatial modes present in the illumination. Different spatial modes produce distinct speckle patterns which sum in intensity, thereby reducing the speckle contrast to $C=M^{-1/2}$, where $M$ is the number of spatial modes [24]. Of course, even if the XLMA fiber ASE was distributed equally among all ~360 passive modes of the fiber, we would naively expect the speckle contrast to be reduced to only $360^{-1/2}$ = 0.05. However, the measured speckle contrast of 0.02 shown in Fig. 2(b) would require contributions from ~2500 mutually incoherent modes if there were no spectral compounding. However, the broadband ASE does enable spectral compounding, as different spectral modes can also contribute to the speckle reduction. Still, it remains difficult to estimate just how many spatial modes are excited from the spectrally integrated speckle patterns measured above. Also the information of the number of spatial modes at any given wavelength is important for applications such as spectral-domain OCT in which a spectrally-resolved detection would limit the effect of spectral compounding.

In order to separate the effect of averaging over the spatial modes from the effect of spectral compounding, we then measured the speckle pattern at individual wavelengths using an imaging spectrometer (Acton Research SpectraPro 300i). To do this, we coupled the emission from the fiber ASE source to a passive, 1 meter long multimode fiber (105μm-diameter core, NA=0.22). The distal end of the multimode fiber was then collimated onto the entrance slit of an imaging spectrometer. At the exit port of the spectrometer a CCD camera (Andor Newton) recorded the spectrally-dispersed one-dimensional (1D) speckle. In the 2D image taken by the CCD camera, the horizontal axis corresponded to wavelength, and the vertical axis to space. In this measurement, the multimode fiber played the role of the diffuser, producing distinct speckle patterns for different spatial modes of the ASE source, while efficiently coupling light into the entrance slit of the spectrometer. However, the speckle patterns formed at the end of a multimode fiber are known to vary as a function of wavelength [26], and a sufficiently long multimode fiber, combined with a broadband light source, can effectively reduce the spatial coherence [27,28]. To confirm that the multimode fiber did not reduce the measured speckle contrast for individual spectral channels resolved by the spectrometer, we first coupled a spatially coherent supercontinuum source (Fianium WhiteLase SC400-4) into the multimode fiber. As shown in Fig. 3(a), the supercontinuum source produced high-contrast speckle in space at each wavelength. Since the spectral correlation of the speckle pattern (corresponding to the spectral correlation of the multimode fiber) was readily resolved by the spectrometer, we know that the passive multimode fiber will not reduce the contrast of spectrally resolved speckle produced by the ASE source.

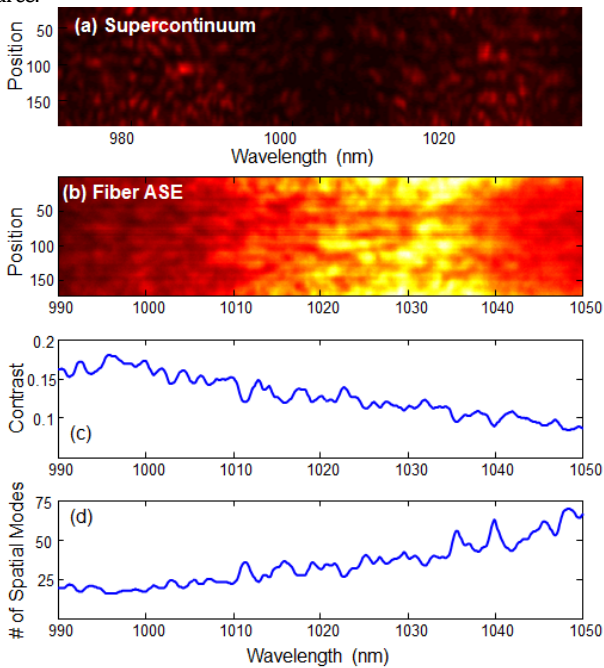

Fig. 3. (a) Spectrally-dispersed speckle pattern formed by coupling emission from a supercontinuum source through a multimode fiber into an imaging spectrometer. The high-contrast speckle observed at each wavelength confirms that the supercontinuum source has high spatial coherence. (b) Spectrally-dispersed speckle pattern formed by the fiber ASE. The reduced speckle contrast at each wavelength is indicative of emission distributed among many spatial modes. (c) The spectrally-resolved speckle contrast is calculated from the 1D speckle pattern in space at each wavelength in (b). (d) The number of spatial modes present at each wavelength is then calculated from the spectrally-resolved speckle contrast, indicating that emission from the fiber ASE source is distributed among 25 to 75 spatial modes.

We then repeated the experiment using the fiber ASE source. The spectrally-dispersed speckle pattern formed by the fiber ASE source is shown in Fig. 3(b). The speckle contrast at any given wavelength is clearly reduced in comparison to the supercontinuum source due to the presence of many spatial modes in the fiber ASE. Based on the image in Fig. 3(b), we then calculated the spectrally-resolved speckle contrast, as shown in Fig. 3(c). From this contrast we then estimated the number of spatial modes present at each wavelength, $M=C^{-2}/2$, where the factor of 2 accounts for polarization mixing in the multimode fiber. Figure 3(d) shows that the number of spatial modes increased with wavelength from ~25 at $\lambda$ = 990 nm to almost 75 spatial modes at $\lambda$ = 1050 nm, near the peak of the emission spectrum. Note that due to the responsivity of the Si CCD camera used to record the spectrally-dispersed speckle patterns, we were unable to measure the number of modes present in the long-wavelength half of the emission spectrum. Nonetheless, Fig. 3(d) illustrates that the number of spatial modes increases from the tail to the center of the gain spectrum. We also note that although the XLMA fiber supports ~360 passive modes, the emission was distributed among less than one fourth of these modes, even at the peak of the gain spectrum. This could be the result of increased bending loss experienced by the higher order modes and/or the mode competition for gain. Nonetheless, the XLMA fiber ASE source distributed emission among many more spatial modes than the multimode SLD, enabling efficient speckle suppression where the SLD did not.

It was initially surprising to find that the multimode fiber ASE source supported such a large number of spatial modes, whereas the ASE produced by a semiconductor-based multimode SLD maintained relatively high spatial coherence and produced high-contrast speckle. In addition, we previously observed similarly strong mode competition for gain in a semiconductor-based, multimode Fabry-Perot laser in which lasing occurred in only a few of the ~450 transverse spatial modes supported by the cavity [13]. However, we believe there are at least two factors which facilitate the fiber ASE source to support a large number of modes. First, the SLD and the multimode Fabry-Perot laser both used semiconductor quantum wells as gain materials which allow for efficient carrier diffusion, thereby increasing the effects of mode competition [29]. In contrast to a quantum well, the Yb dopants in the XLMA fiber are spatially localized, leading to spatial hole burning which can reduce the effects of mode competition [30,31]. Second, the fiber bending and imperfections (local fluctuation of the refractive index and variation of fiber cross section) in the XLMA fiber introduce mode coupling such that a mode which initially experiences strong gain may couple into a mode with lower gain, thereby equalizing the gain experienced by different modes over the length of the fiber. The reduction of mode-dependent gain in the multimode fiber favors the many mode operation [32].

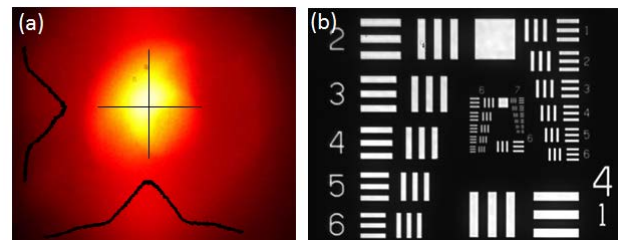

Fig. 4. (a) The spatial profile of the output beam from the fiber ASE source, measured by a CCD camera placed in the path of the collimated beam, showing a uniform profile with Gaussian cross section and a divergence angle of less than 6°. (b) An image of a U.S. Air Force resolution chart illuminated in transmission through a static ground glass by the fiber ASE source, which is free of speckle.

In addition to the low spatial coherence, the temporal coherence of the fiber ASE source is also low, which is well suited for ranging applications such as optical coherence tomography (OCT) or frequency-resolved coherent LiDAR [19]. For example, the 3dB bandwidth of 74 nm would provide an axial resolution of 6.6 μm in OCT. In addition to providing both low spatial and low temporal coherence, the fiber ASE source also exhibits high directionality, with a divergence angle of less than 6° (dictated by the NA = 0.1 of the fiber).

Despite the participation of many spatial modes, the spatial profile of the output beam from the fiber ASE source is smooth and well suited for illumination in imaging applications, as confirmed by the image of the collimated beam in Fig. 4(a). Finally, we used the fiber ASE as an illumination source to image a U.S. Air Force resolution chart through a static ground glass in transmission mode. A speckle-free full-field image was obtained as shown in Fig. 4(b).

While the fiber ASE source provides speckle-free illumination similar to a thermal source or LED, it also produces much higher power per mode, which could enable high-speed imaging or illumination at large distances. As a quantitative comparison, we calculated the photon degeneracy which describes the number of photons per coherence volume. The photon degeneracy parameter $\delta = (P\,\delta z)/(h\nu\,c\,M)$, where $P$ is the emission power, $\delta z$ is the temporal coherence length, $h\nu$ is the photon energy, $c$ is the speed of light, and $M$ is the number of spatial modes [33]. Based on our measurement shown in Fig. 3, revealing that the fiber ASE is distributed on average among ~50 spatial modes, we calculated a photon degeneracy of $\delta \sim 600$. This is more than five orders of magnitude higher than the photon degeneracy of a thermal source (e.g. at 4000K temperature, $\delta \sim 10^{-3}$ [33]), and more than four orders of magnitude higher than that of a bright LED ($\delta \sim 10^{-2}$). It is also competitive with recent a demonstration of relatively narrowband, low-spatial-coherence chaotic microcavity laser ($\delta \sim 100$) [13]. In addition, the high-power commercial SLD shown to produce speckle of contrast ~0.2 in Fig. 2(d) exhibits similar degeneracy of $\delta \sim 600$, despite maintaining relatively high spatial coherence and a low number of spatial modes.

In summary, we have demonstrated a fiber ASE source which combines high power per mode with low spatial and low temporal coherence. The ASE source provides 270 mW of CW emission with 74 nm 3dB bandwidth centered at $\lambda = 1055$ nm. We characterized the spatial coherence and found that the emission is distributed among as many as ~70 spatial modes. A further increase of the number of spatial modes is possible by increasing gain or using a fiber with a larger gain core. The emission exhibits a small divergence angle and uniform spatial profile, making it well-suited as an illumination source in full-field imaging and ranging applications. In the future, we expect the pump efficiency and output power of the fiber ASE source will be dramatically improved by matching the geometry of the passive multimode fiber to the pump core of the XLMA fiber.

National Institutes of Health (NIH) Grant No. 1R21EB016163-01A1 and 1R21HL125125-01A1, Office of Naval Research Grant No. ONR MURI SP0001135605

We thank Yaron Bromberg for useful discussions.

## References

1. P. Wang, J. K. Sahu, and W. A. Clarkson, Opt. Lett. **31**, 3116–3118 (2006).
2. M. Rossetti, J. Napierala, N. Matuschek, U. Achatz, M. Duelk, C. Vélez, A. Castiglia, N. Grandjean, J. Dorsaz, and E. Feltin, Proc. SPIE **8252**, 825208 (2012).
3. W. Denzer, M. L. Hamilton, G. Hancock, M. Islam, C. E. Langley, R. Peverall, and G. A. D. Ritchie, Analyst **134**, 2220 (2009).
4. A. F. Fercher, W. Drexler, C. K. Hitzenberger, and T. Lasser, Reports Prog. Phys. **66**, 239 (2003).
5. H. S. Choi, H. F. Taylor, and C. E. Lee, Opt. Lett. **22**, 1814 (1997).
6. B. Lee, Opt. Fiber Technol. **9**, 57 (2003).
7. B. Karamata, M. Leutenegger, M. Laubscher, S. Bourquin, T. Lasser, and P. Lambelet, J. Opt. Soc. Am. A **22**, 1380 (2005).
8. B. Redding, M. A. Choma, and H. Cao, Nat. Photonics **6**, 355 (2012).
9. A. Mermillod-Blondin, H. Mentzel, and A. Rosenfeld, Opt. Lett. **38**, 4112 (2013).
10. B. H. Hokr, M. S. Schmidt, J. N. Bixler, P. N. Dyer, G. D. Noojin, B. Redding, R. J. Thomas, B. A. Rockwell, H. Cao, V. V. Yakovlev, M. O. Scully, "A narrow-band speckle-free light source via random Raman lasing," *(In Review)* ArXiv: 1505.07156 (2015).
11. M. Nixon, B. Redding, A. A. Friesem, H. Cao, and N. Davidson, Opt. Lett. **38**, 3858 (2013).
12. R. Chriki, M. Nixon, V. Pal, C. Tradonsky, G. Barach, A. a. Friesem, and N. Davidson, *"Manipulating the spatial coherence of a laser source,"* Opt. Express **23**, 12989 (2015).
13. B. Redding, A. Cerjan, X. Huang, M. Larry, A. D. Stone, and M. A. Choma, Proc. Natl. Acad. Sci. **112**, 1304 (2015).
14. F. Riechert, G. Verschaffelt, M. Peeters, G. Bastian, U. Lemmer, and I. Fischer, Opt. Commun. **281**, 4424 (2008).
15. G. Craggs, G. Verschaffelt, S. K. Mandre, H. Thienpont, and I. Fischer, IEEE J. Sel. Top. Quantum Electron. **15**, 555 (2009).
16. G. Verschaffelt, G. Craggs, M. L. F. Peeters, S. K. Mandre, H. Thienpont, and I. Fischer, IEEE J. Quantum Electron. **45**, 249 (2009).
17. J.-F. Seurin, G. Xu, V. Khalfin, A. Miglo, J. D. Wynn, P. Pradhan, C. L. Ghosh, and L. A. D'Asaro, "Progress in high-power high-efficiency VCSEL arrays," *Proc. SPIE* **7229**, 722903 (2009).
18. W. Drexler and J. G. Fujimoto, *Optical Coherence Tomography* (Springer-Verlag, Berlin Heidelberg, 2008).
19. W. C. Swann and N. R. Newbury, Opt. Lett. **31**, 826 (2006).
20. A. Langner, G. Schötz, M. Such, T. Kayser, V. Reichel, S. Grimm, J. Kirchhof, V. Krause, and G. Rehmann, Proc. SPIE **6873**, 687311 (2008).
21. K. Okamoto, *Fundamentals of Optical Waveguides* (Elsevier, 2006).
22. D. Marcuse, Appl. Opt. **14**, 3016 (1975).
23. N. Savage, Nat. Photon. **3**, 114 (2009).
24. J. W. Goodman, *Speckle Phenomena in Optics* (Roberts & Company, 2007).
25. S. Roelandt, Y. Meuret, A. Jacobs, K. Willaert, P. Janssens, H. Thienpont, and G. Verschaffelt, "Human speckle perception threshold for still images from a laser projection system," *Opt. Express* **22**, 23965 (2014).
26. B. Crosignani, B. Daino, and P. Di Porto, J. Opt. Soc. Am. **66**, 1312 (1976).
27. N. Takai and T. Asakura, J. Opt. Soc. Am. A **2**, 1282 (1985).
28. J. Kim, D. T. Miller, E. Kim, S. Oh, J. Oh, and T. E. Milner, J. Biomed. Opt. **10**, 064034 (2005).
29. H. Statz, C. L. Tang, and J. M. Lavine, J. Appl. Phys. **35**, 2581 (1964).
30. J. Hao, P. Yan, and M. Gong, Opt. Commun. **287**, 167 (2013).
31. A. Cerjan, Y. Chong, and A. D. Stone, Opt. Express **23**, 6455 (2015).
32. K.-P. Ho and J. M. Kahn, J. Light. Technol. **32**, 614 (2014).
33. L. Mandel and E. Wolf, *Optical Coherence and Quantum Optics* (Cambridge University Press, 1995).